\documentclass{eas}
\usepackage{graphicx}
\usepackage{amssymb,amsmath}
\usepackage{natbib}
\begin{document}

\title{Evolution of PAHs in protoplanetary disks} 
\author{Inga Kamp}\address{Kapteyn Astronomical Institute, University of Groningen, PO Box 9513, 2300 AV Groningen, The Netherlands}
\begin{abstract}
Depending on whom you ask, PAHs are either the smallest dust particles or the largest gas-phase molecules in space. Whether referred to as gas or dust, these PAHs can contain up to ~20\% of the total cosmic carbon abundance and as such also play an important role in the carbon chemistry of protoplanetary disks. The interpretation of PAH bands is often a complex procedure involving not only gas physics to determine their ionization stage and temperature, but also radiative transfer effects that can bury these bands in a strong thermal continuum from a population of larger dust particles. 

PAHs are most readily seen in the spectral energy distributions (SEDs) of disks around Herbig AeBe stars where they are photoprocessed by the stellar radiation field. Resolved images taken in the PAH bands confirm their origin in the flaring surfaces of circumstellar disks: if the SED is consistent with a flat disk structure (less illuminated), there is little or no evidence of PAH emission. The very low detection rates in the disks around T Tauri stars often require an overall lower abundance of PAHs in these disk surface as compared to that in molecular clouds. 
 
In this review, I will adress three aspects of PAHs in protoplanetary disks: (1) Do PAHs form in protoplanetary disks or do they originate from the precursor molecular cloud? (2) Is the presence of PAH features in SEDs a consequence of the disk structure or do PAHs in fact shape the disk structure? (3) How can we use PAHs as tracers of processes in protoplanetary disks?
\end{abstract}
\maketitle
\section{Introduction}

Polycyclic aromatic hydrocarbons (PAHs) are considered amongst the smallest dust grains even though their sizes range from macromolecules with $10$ carbon atoms, such as naphthalene to large complexes with more than 100 carbon atoms. They are identified mainly through their near- and mid-IR bands in very different astrophysical environments such as the interstellar medium (ISM), disks around young stars, and the circumstellar environment of AGB stars. I will focus in the following on the role of PAHs in star formation and especially on their relevance in protoplanetary disks.

Protoplanetary disks are complex in nature, because they span a wide range of physical conditions, i.e. density and temperature (and irradiation), but also show orders of magnitude different dynamical timescales between the inner and outer disk. Gas and dust in these disks are not necessarily coupled and co-spatial. And a number of processes that operate in protoplanetary disks such as dust coagulation and settling, gas dispersal, and planet formation affect gas and dust often in different ways. 

Their nature positions PAHs in terms of their properties somewhere between large gas molecules and small dust grains. Due to their small size, they efficiently couple to the widespread gas in protoplanetary disks and are thus able to escape settling processes that affect the much larger dust grains. Due to their low ionization potentials, PAHs are easily ionized by the stellar UV photons and the ejected electrons can efficiently heat the surrounding gas (photoelectric heating). If they are abundant (23\% of carbon dust in the form of PAHs), they can also present an important opacity source and absorb up to 40\% of the total radiation \citep{habart2004}. Often the fraction of carbon dust in PAHs is assumed to be much smaller in disks, e.g.\ less than 10\% \citep{geers2006, dullemond2007}. PAHs are stochastically heated particles and undergo rapid and extreme temperature fluctuations that lead to their emission in the aromatic infrared bands (AIBs).

I start out with a brief summary of PAH observations in protoplanetary disks and their detection frequency (Sect.~\ref{Obs}). Sect.~\ref{disk:em} provides a short overview of the PAH excitation mechanism and radiative transfer models. The next two sections then discuss the role of PAHs in disks chemistry (Sect.~\ref{disk:chem}) and physics (Sect.~\ref{disk:phys}) in more detail. The last section will then address three questions concerning PAH evolution in disks: (1) Do PAHs form in protoplanetary disks or do they originate from the precursor molecular cloud? (2) Is the presence of PAH features in SEDs a consequence of the disk structure or do PAHs in fact shape the disk structure? (3) How can we use PAHs as tracers of processes in protoplanetary disks?

\section{Some observational facts}
\label{Obs}



PAHs have been rarely observed in embedded low mass protostars. However, as the sources evolve into class\,{\sc ii} and class\,{\sc iii} objects, detection rates increase and there is a clear dependency on the UV radiation field of the central star. PAHs are observed in 57\% of the Herbig disks \citep{acke2004}, but only in 8\% of the T Tauri disks \citep{geers2006}. Various studies \citep{sloan2007, boersma2008, keller2008} noted clear changes in the feature shape and shifts of the central wavelength, which they attribute to signatures of PAH processing in the young star's environment (see Acke, elsewhere this volume, for a more detailed discussion). It seems that the wealth of observational data especially from Spitzer available now for several objects (see below) warrants a much more detailed understanding of the disk structure.
For example, \citet{fedele2008} find the PAH emission coming from the upper gas surface in the disk around HD101412; the PAHs are co-spatial with the [O\,{\sc i}]\,6300~\AA\ emission of the gas, which is generally thought to be a by-product of the photodissociation of OH in the UV irradiated disk surface. On the other hand, the SED and near-IR visibilities show that the larger dust grains are not directly exposed to the stellar radiation and must reside at lower heights in the disk. This lends support to the idea that PAHs stay well mixed with the gas. Along the same lines, \citet{verhoeff2010} find that the disk around HD95881 consists of a thick puffed up inner rim and an outer region in which the gas still has a flaring structure, while larger dust grains have settled to the midplane. Such systems are thought to be in a transitional stage between a gas-rich flaring and gas-poor shadowed configuration.


\section{PAH emission from disks}
\label{disk:em}

\begin{figure}[thbp]
\includegraphics[scale=0.34]{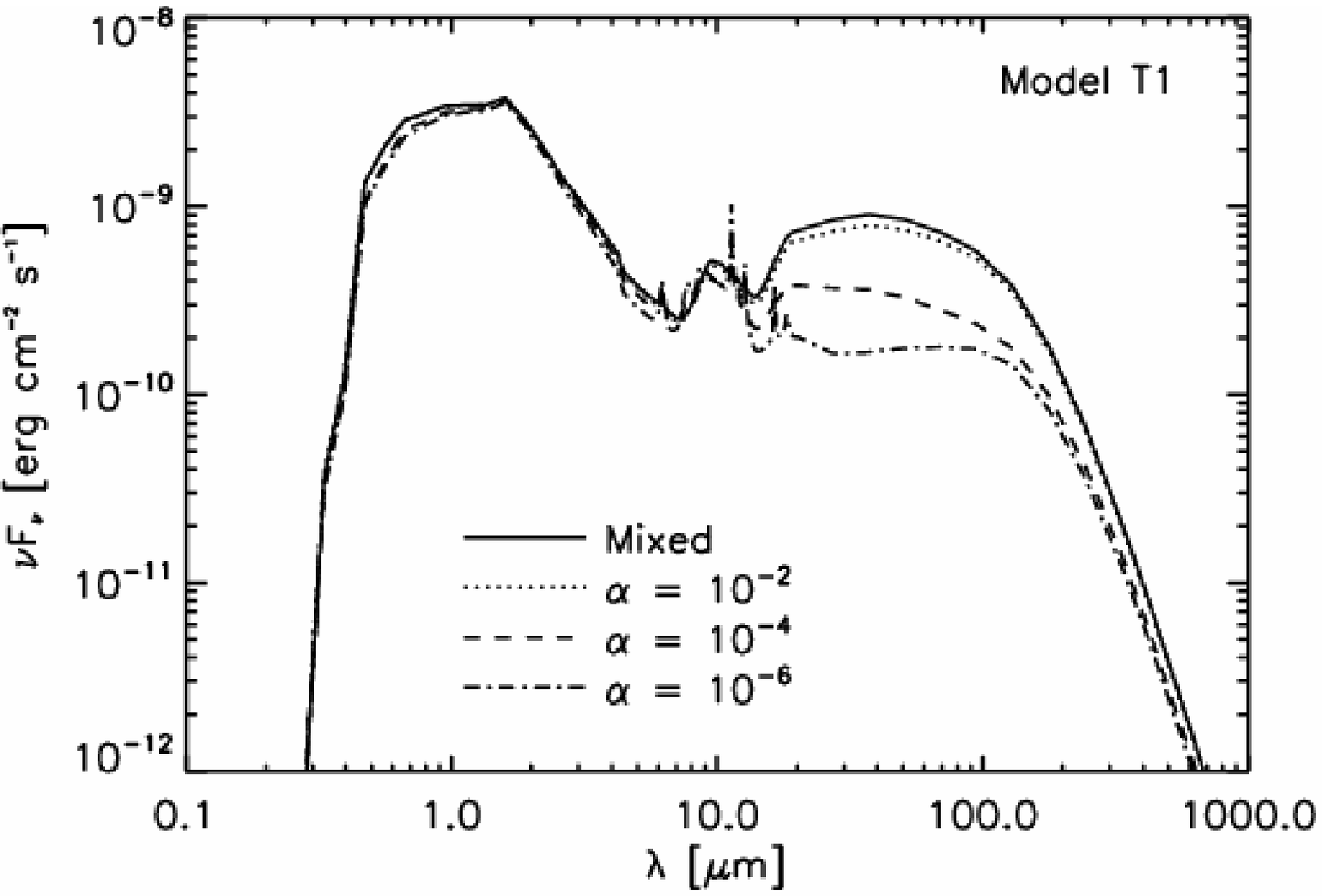}
\includegraphics[scale=0.34]{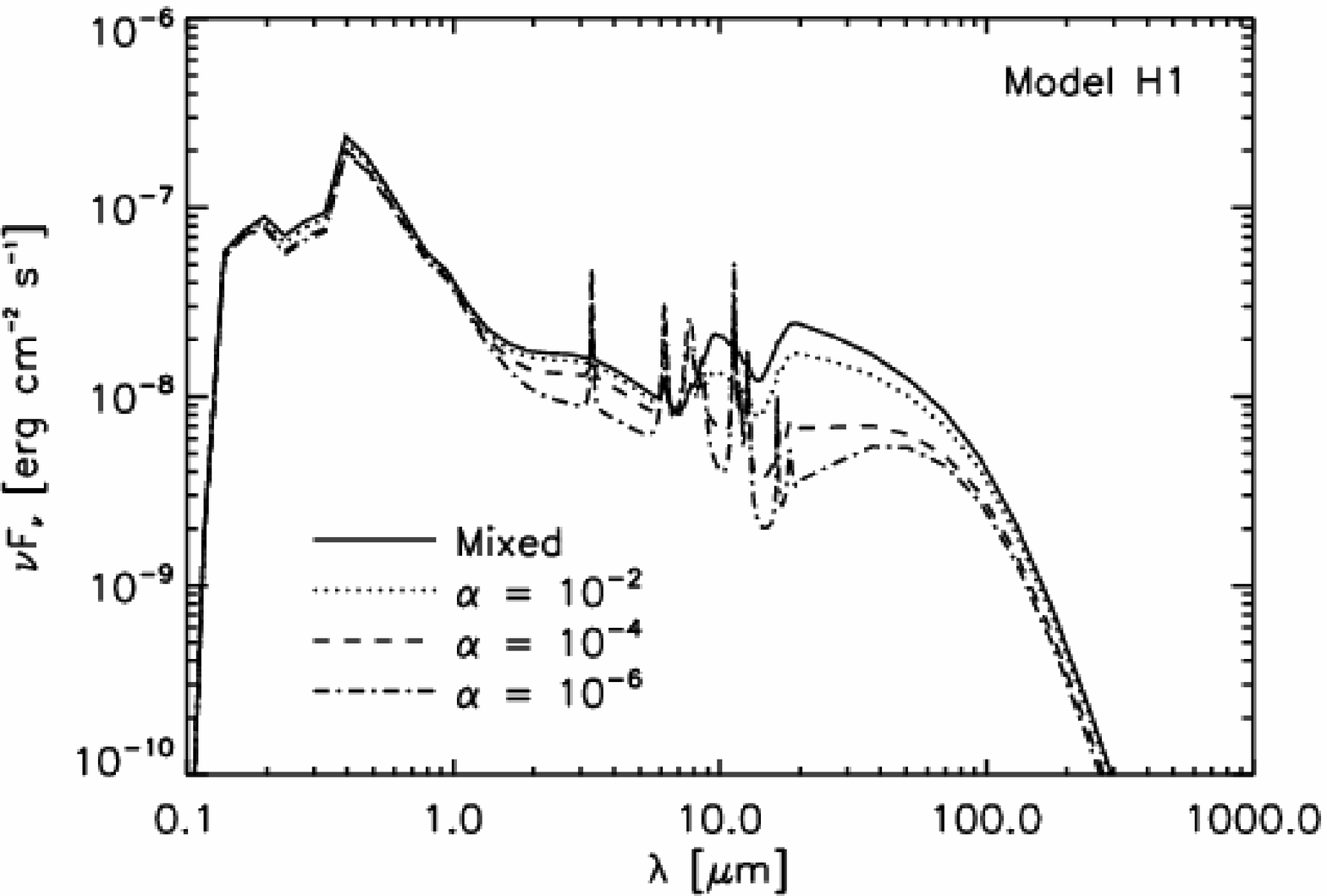}
\vspace*{-8mm}
\caption{The SEDs of a T Tauri disk model (Model T1) and a Herbig disk model (Model H1) at an evolutionary age of $10^6$~yr (inclination of 45 degrees, $d=100$~pc). The different curves show the SED before sedimentation (solid), and after sedimentation ($t=10^6$~yr) for several values of the disk viscosity parameter: $\alpha= 10^{-2}$ (dotted) , $\alpha= 10^{-4}$ (dashed) and $\alpha= 10^{-6}$ (dot-dashed). The figure is taken from \citet{dullemond2007}.}
\vspace*{-3mm}
\label{dullemond2007_fig2}
\end{figure}

A detailed understanding of the PAH emission mechanisms is essential to judge the diagnostic power and impact of PAHs on the disk structure. In the following, the problem is split into the excitation mechanism (ionization state, single versus multi-photon excitation) and the radiative transfer in the presence of a strong dust continuum and competing features such as the silicate emission at 10~$\mu$m. The combination of these two aspects leads to  differences in the spatial appearance of the various PAH emission features (see e.g.\ Fig.~\ref{dullemond2007_fig2}).

Studies of galactic emission have shown that PAHs contain up to a few percent of the total carbon dust mass \citep[e.g.][]{pendleton1994}; the typical PAH carbon abundance found from galactic PDRs and ISM is [C/H]$_{\rm PAH} \approx 5 \times 10^{-5}$ \citep{boulanger1988, habart2001,desert1990,dwek1997,li2001}. For such an extreme case of PAH abundance in disks, \citet{habart2004} find that PAHs reprocess $\sim 30$\% of the impinging stellar radiation (which is $\sim 0.3$\% of the total luminosity in the case of a flaring disk surface) in a fiducial Herbig disk model ($0.3-300$~AU, $0.1$~M$_\odot$, M$_\ast = 2.4$~M$\odot$, L$_\ast = 32$~L$_\odot$); since PAHs then present a significant fraction of the total dust opacity, their presence has a visible effect on the overall SED (see their Figs.~1 and 3). \citet{habart2004} find that $\sim 25$\% of the flux that PAHs absorb in the inner disk is consumed by their evaporation. \citet{geers2006} found from an extensive comparison of observed PAH features ({\it Spitzer} spectra) and detailed radiative transfer disk models that the typical PAH abundance in disks must be 10-100 times lower than that in the ISM. Hence, the total dust opacity should be  hardly affected by PAHs.

\subsection{The excitation mechanism of PAHs}

The intensity emitted in a specific AIB roughly scales with the incident FUV radiation provided that the charge state is held constant \citep{li2003, habart2004}. This directly reflects the single photon excitation mechanism (transiently heated grains) as opposed to the radiative equilibrium that holds for larger grains. Hence, for PAHs one generally defines a temperature distribution function ${\rm d}P/{\rm d}T$ that describes the probability of finding a PAH in a particular temperature interval. Due to the single photon excitation mechanism, this temperature distribution depends not on the strength, but on the color of the radiation field. The different PAH bands then require different excitation temperatures, with the short wavelength bands (C-H, C-C stretching modes) requiring higher temperatures than the long wavelength bands (bending modes). 

However, spinning of charged dust grains at thermal rates is an alternative excitation mechanism that works at much longer wavelengths \citet{rafikov2006} (see Draine elsewhere this volume). This microwave emission would trace small dust grains in general (VSGs and PAHs) and it has the potential to trace PAHs even down to the midplane.


\subsection{Feature-to-continuum ratios}

\citet{manske1999} performed a two-dimensional radiative transfer calculation for disk plus envelope systems. They include various grain populations, amongst them also the  transiently heated VSGs and PAHs. One of their main results is that the PAH features become weaker with increasing mass fraction of VSGs. However, it proved difficult to entirely surpress the PAHs in the SEDs from Herbig stars with disks and envelopes. The later models by \citet{dullemond2007} confirm that the VSGs compete with PAHs for the stellar UV photons, thus decreasing the PAH band strength. The SED on the other hand changes only slightly between 20 and 30~$\mu$m where VSGs produce a bump in the disk emission.

\subsection{Effects of ionization, dehydrogenation and size}

The various AIBs are attributed to vibrations of different bonds in the PAH molecule and their strength changes with the PAH ionization, dehydrogenation state, and size. The following paragraphs summarize our understanding of PAH emission based on results from \citet{habart2004}. 

The $3.3~\mu$m band is a C-H stretching mode, while the 6.2 and $7.7~\mu$m bands originate from C-C stretching modes. The C-H in-plane and out-of-plane bending modes cause the $8.6~\mu$m band and the $11.3$, $11.9$, and $12.7~\mu$m features respectively. 

The strong effect that the ionization state (neutral or cation) has on the relative PAH band intensities is known from theory and experiment (see references in Pauzat and Oomens, elsewhere in this volume).
The 6.2 and $7.7~\mu$m bands are stronger in ionized PAHs; the $8.6~\mu$m band is stronger in cations than in neutrals. On the contrary, the 11.3, 11.9, $12.7~\mu$m bands decrease for charged PAHs. Otherwise, PAH anion band strength is often between that of neutrals and cations.

With strong dehydrogenation, the C-H features would disappear. This is only relevant for the  smallest PAHs ($N_{\rm C} \leq 25$) \citep{tielens1987, allain1996}. Theory predicts that the C-H features (3.3, 8.6, 11.3, 12.7) become weaker, while the C-C features (6.2, 7.7) become slightly stronger.

In addition to charge and hydrogenation state, the size of the PAH also impacts the excitation state. Small PAHs favor the $3.3~\mu$m band, while larger ones emit preferentially at longer wavelengths. This leads to a degeneracy between charge and size.

\subsection{The spatial origin of various PAH features}

\citet{habart2006} spatially resolved the 3.3~$\mu$m PAH feature using NAOS-CONICA at the VLT (AO providing $\sim 0.1"$ resolution). The emission originates from within 30~AU of the star and is significantly more extended than the adjacent continuum. This is in agreement with the theoretical models that find the higher excitation temperatures required for the $3.3~\mu$m feature mostly in the inner disk \citep[$r<50$~AU;][]{habart2004}. \citet{geers2007a} and \citet{geers2007b} also report that the disk emission in some targets is more extended within the PAH bands than within the adjacent continuum. Just as a note of caution, the emission from VSGs can also be more extended than that of large dust grains.

\section{PAH chemistry in disks}
\label{disk:chem}

As large gas molecules, PAHs take part in the chemical reaction networks of protoplanetary disks. They can undergo ionization (photoelectric effect), electron recombination and attachment, charge exchange, photodissociation with C or H loss, and H addition reactions. Trapping of PAHs onto dust grains or within ices \citep{gudipati2003} are more relevant in the cold dense midplane that does not contribute to the PAH emission features. Reaction rates with heavier elements can be comparable to those of H, H$_2$, but often the abundances of those heavier elements are much lower. However, in some regions of protoplanetary disks, where H is efficiently converted into water and/or OH, reactions with O and OH can play an important role in breaking PAHs down \citep{kress2010}.

The convention often found in the literature \citep[e.g.][]{lepage2001} is to characterize PAHs by the number of C atoms $N_C$. The normal number of hydrogen atoms, $N_H^0$ is the number of hydrogen atoms when one hydrogen atom is attached to each peripheral carbon atom which is bonded to exactly two other carbon atoms. In the following, PAH stands representative for any size PAH and the different charge states are indicated by suerscripts, e.g. PAH$^+$. A recent review on PAH chemistry is given by \citet{tielens2008} and the formation of PAHs is discussed by Cherchneff (this volume).

\subsection{PAH hydrogenation and dissociation}

The stability and hydrogenation of PAHs is mainly determined by the strength of the UV environment ($\lambda < 2000$~\AA) in which they reside \citep[the ionization potential of PAHs is typically ${\rm IP} > 6$~eV, see][]{weingartner2001}. While photons of 3-10 eV can lead to the loss of H (3.2 eV), C$_2$H$_2$ (4.2 eV), C (7.5 eV) and C$_2$ (9.5 eV), X-ray photons are more energetic and can destroy entire PAHs with $N_{\rm C}=100$ \citep{siebenmorgen2009}.

Small PAHs ($N_{\rm C}=24$) are easily destroyed in protoplanetary disks on short timescales ($< 3$~Myr). Larger PAHs survive more easily in the irradiation environment of protoplanetary disks. PAHs with $N_{\rm C}=96$ for example survive down to 5~AU in the disk around a Herbig star, while smaller ones ($N_{\rm C}=50$) are efficiently destroyed out to a few 10 AU \citep[see Fig.7 of][]{visser2007}. The weaker radiation field of T Tauri stars is not capable of destroying PAHs unless these stars have a significant UV excess. X-ray and EUV photons are so efficient that they destroy PAHs wherever they can reach them \citep{siebenmorgen2009}.

\subsection{Charge distribution of PAHs in disks}

\begin{figure}[htbp]
\includegraphics[scale=0.7]{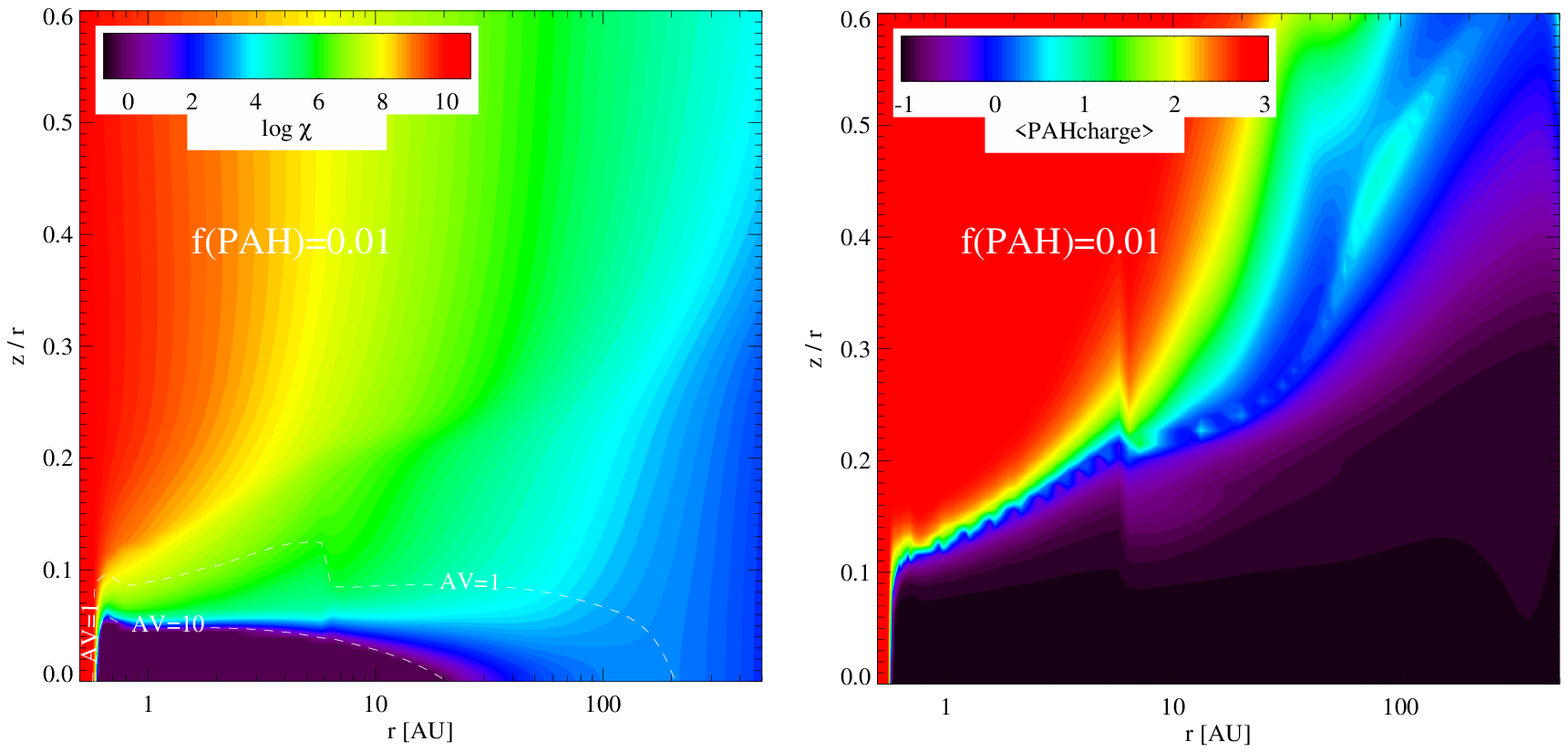}
\vspace*{-8mm}
\caption{{\bf Left:} Strength of the UV radiation field $\chi$ in the protoplanetary disk around a typical Herbig star ($M_\ast = 2.5$~M$_\odot$, $T_{\rm eff} = 10000$~K). The main parameters are: $M_{\rm disk}=0.025$~M$_\odot$, a surface density power law profile $\epsilon=1.0$, a grain size distribution between $a_{\rm min}=0.05~\mu$m and $a_{\rm max}=1$~mm with a power law exponent of $3.5$, and a dust-to-gas mass ratio of 0.01, a PAH abundance $f_{\rm PAH}$ of 1\% relative to the ISM. $\chi$ is defined as the integral of the radiation field between 912 and 2050~\AA\ normalized to that of the Draine field \citep{draine1996}. {\bf Right:} Charge distribution of PAHs. Note that regions where PAHs carry multiple positive charges have low particle densities (see Fig.~\ref{PAH_diskstructure}).}
\label{PAH_charges}
\vspace*{3mm}
\includegraphics[scale=0.7]{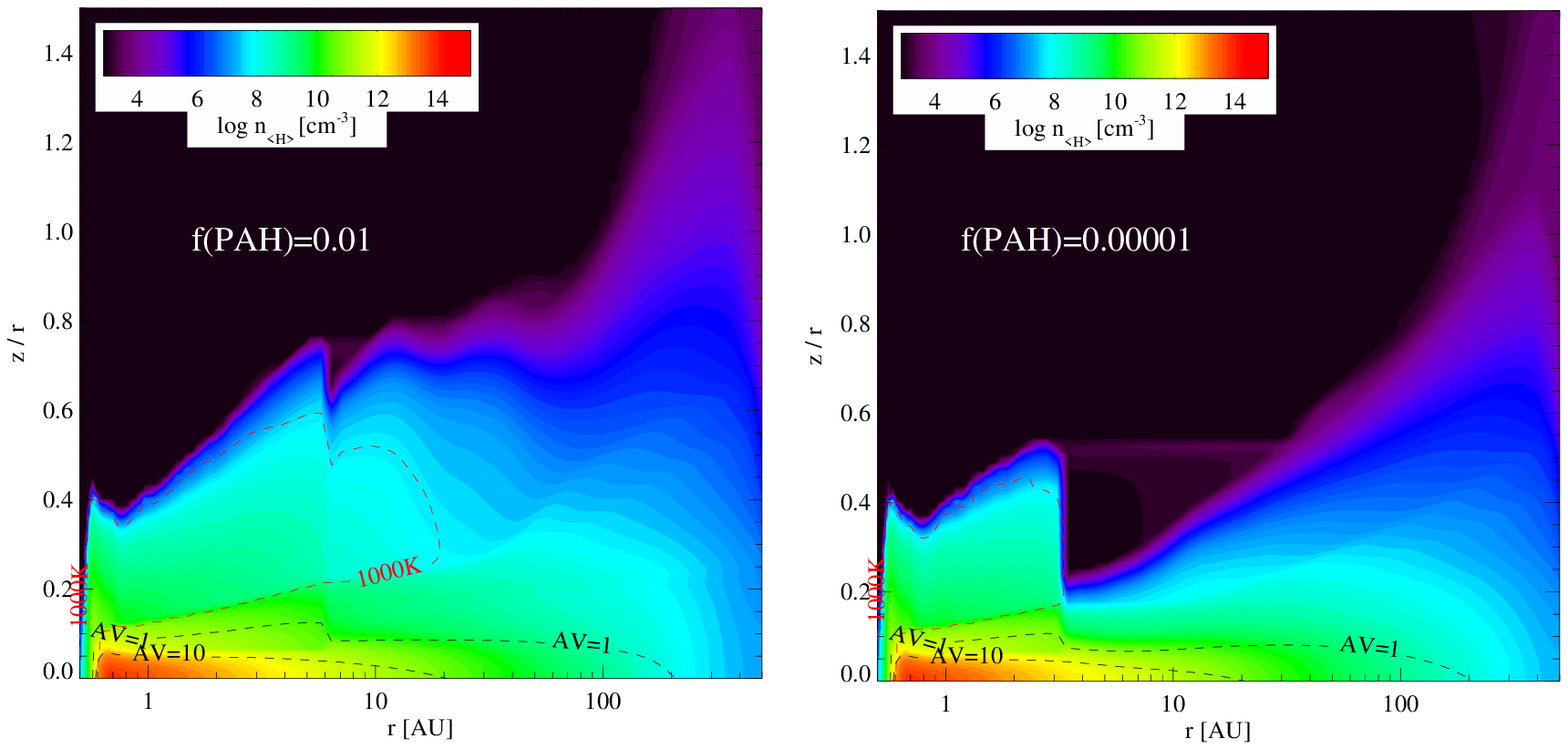}
\vspace*{-8mm}
\caption{Density structure of a protoplanetary disk around a 2.5~M$_\odot$ Herbig star ($T_{\rm eff} = 10000$~K). The main parameters are: $M_{\rm disk}=0.025$~M$_\odot$, a surface density power law profile $\epsilon=1.0$, a grain size distribution between $a_{\rm min}=0.05~\mu$m and $a_{\rm max}=1$~mm with a power law exponent of $3.5$, and a dust-to-gas mass ratio of 0.01. The left figure shows a model with a PAH abundance relative to the ISM of $0.01$ and the right figure one of $10^{-5}$. The black dashed lines show the location of the $A_V=1$ and 10 surfaces, while the red contour line indicates the location of the hot gas ($T_{\rm gas} > 1000$~K)}
\label{PAH_diskstructure}
\end{figure}

The PAH charge state in disks generally has a layered structure with cations being on top, followed by neutrals and anions towards the dense shielded midplane \citep{visser2007}.  Fig.~\ref{PAH_charges} shows that PAHs get easily double ionized (PAH$^{2+}$) in the surfaces of disks around Herbig stars due to the strong stellar UV radiation field (detailed model description in Sect.~\ref{disk:phys}). This result depends strongly on the size of the PAH. The disk model shown here uses circumcoronene ($N_{\rm C}=54$, $N_{\rm H}=18$, IP$(1)=6.243$, IP$(2)=9.384$, IP$(3)=12.524$). Due to the weak UV radiation field of T Tauri stars, disks around low mass stars show mostly neutral PAHs and anions \citep[$\sim 50\%$ of the entire population,][]{visser2007}. 


\subsection{H$_2$ formation on PAHs}

\citet{jonkheid2006} consider that for evolved disks (HD\,141569), H$_2$ might form more efficiently in reactions with PAHs than on the dust grains in the disk ($1~\mu{\rm m} < a < 1$~cm). Due to the lack of small grains, the total dust surface area available from such a distribution is small compared to the typical ISM size distribution and the H$_2$ formation rate depends directly on the total available surface area.

\section{The role of PAHs in disks physics}
\label{disk:phys}

PAHs are not relevant for the dust energy balance, i.e.\ they do not present a significant fraction of the dust opacity. However, as large molecules, they efficiently absorb UV radiation and get photoionized. The energetic electrons released in this way efficiently heat the surrounding gas and determine its temperature. Thus the vertical disk structure --- as set by the gas temperature --- depends sensitively on this heating source.


\subsection{The vertical disk structure}

The thermo-chemical disk code {\sc ProDiMo} calculates the vertical hydrostatic disk structure, the chemical composition, and gas \& dust temperatures self-consistently. The basic physics and chemistry of the code is summarized in \citet{woitke2009} while additions and new features are described in \citet{kamp2010} (UV radiative transfer and UV-photochemical rates) and Woitke et al.\ (in preparation) (UV pumping, PAH heating and ray tracing of gas lines). 

The models shown here (van der Plas et al.\ in preparation) use a chemical network consisting of 10 elements, 76 species and 973 reactions, among them also the five ice species: CO, CO$_2$, H$_2$O, CH$_4$, and NH$_3$. The inner disk has 'soft edges', i.e. soft density gradients at the inner and outer radius. 

Fig.~\ref{PAH_diskstructure} shows two models of a disk around a Herbig star that differ only in their fractional PAH abundance. The inner disk wall is directly illuminated by the star while the gas behind it is irradiated under a shallow angle; hence, the inner rim has a higher gas temperature than the disk behind it. Under the assumption of vertical hydrostatic equilibrium, this translates into a 'puffed-up' inner rim. If the rim has a substantial optical depth, it can cast a shadow on the material behind it and thus cause a 'shadow', i.e. a flat disk structure \citep{dullemond2007b}.

The amount of shadowing depends crucially on how efficient the gas is heated by the stellar radiation field. The gas and dust temperatures in these models de-couple in the surface layers down to $A_V \sim 1$. For large gas heating efficiencies, it is thus more difficult to cast a shadow. One of the strongest heating processes in the disk surface is photoelectric heating by PAHs. Hence, the disk with very little PAHs has a much lower heating efficiency. The main heating processes in this case are the photoelectric effect on dust grains and the H$_2$ formation heating. The overall disk structure is cooler and hence much flatter (see e.g.\ the $A_V=1$ contour lines). In addition, the inner rim now casts a very strong shadow beyond 3~AU.

\subsection{PAHs and dust settling}

\citet{dullemond2007} have studied the impact of grain size sorting with height due to dust sedimentation and turbulence equilibrium. They find that the sedimentation enhances the dust features originating from the smallest grains (Fig.~\ref{dullemond2007_fig2}). Even though PAHs are potentially destroyed in the irradiated disk surface, rapid turbulent mixing ensures a continuous supply from deeper layers. 

As a consequence, the inclusion of sedimentation lowers the required PAH abundances to produce a certain feature-to-continuum ratio. The impact on the SED is twofold: (a) sedimentation supresses the continuum by letting small grains that are efficiently heated sink down (b) sedimentation lowers the optical depth in surface layers, thereby letting all UV flux be absorbed by PAHs in surface and thus enhancing the feature strength.

\section{Conclusions}

After an extensive discussion of the PAH chemistry, the role they play in disk physics and their radiative transfer, this section returns to the initially raised three questions and tries to give an answer based on our current understanding.

\subsection{Do PAHs form in protoplanetary disks?}

The formation of PAHs requires high densities and temperatures ($\sim 1000$~K) and high abundances or organic molecules such as e.g.\ acetylene in a low UV radiation environment. Those conditions could likely exist directly behind the inner rim of a protoplanetary disk. Using various initial abundances of C$_2$H$_2$ and CH$_4$ and CO and $(T,P)$ combinations, \citet{morgan1991} showed that PAH formation can in fact be very efficient in certain temperature/pressure regimes of the early solar nebula. More recently, \citet{woods2007} used a full chemical network (including the formation and destruction of C$_2$H$_2$ and CH$_4$ and CO) to show that in a stationary protoplanetary disk (no mixing), benzene can indeed form inside 3~AU and abundances of other organic species are also high (see Woods, elsewhere this volume). \citet{kress2010} define a soot line for disks, which is the dividing line between the location where carbon compounds readily condense into PAHs at high gas temperatures and that where they rather stay in simpler molecules such as CO, CO$_2$, C$_2$H$_2$, CH$_4$. The remaining question is how these {\it in-situ} formed PAHs are mixed to the large radii where they are observed (up to several 10~AU). 

Alternatively, PAHs could form in shocks. \citet{desch2002} invoke shocks for the melting of chondrules in the inner solar nebula. If the shocks are associated with the global disk structure e.g.\ because of spiral density enhancements that travel at a lower pattern speed compared to the gas in which they are embedded, those shocks could be recurring on an orbital timescale. The conditions (e.g.\ temperatures and densities) could then be very similar to those in the outer envelopes of AGB stars where PAHs are known to form efficiently \citep[][Cherchneff elsewhere this volume]{cherchneff1992}.

Important for the question whether PAHs can survive from an earlier ISM phase is the destruction timescale. PAH destruction is mostly driven by UV and X-ray radiation leading to dehydrogenation and photodestruction of the carbon skeleton. In most cases, UV photons can break the weaker C-H bonds thus leading to dehydrogenation. If PAHs absorb more than 21~eV in an interval shorter than their cooling time, the carbon skeleton breaks up, thereby destroying the PAH \citep{guhathakurta1989}. These processes will be most relevant in the surfaces of disks that are exposed to direct UV and X-ray radiation. Another important process is the collisional destruction of PAHs (involving H, O and OH) at high temperatures and pressure ($T \sim 1000$~K, $P \sim 10^{-6}$~bar). \citet{kress2010} find that the overall destruction timescale of PAHs is set by the timescale of the most stable one, since high C$_2$H$_2$ abundances are used to convert smaller PAHs back into bigger ones. 

Something to keep in mind is that the small dust in the form of VSGs and PAHs presents a very distinct grain population in the ISM. The fact that they do not follow the general power law size distribution \citep{draine2004} could indicate that they experience very different processing as compared to the bulk silicate and graphite dust. An alternative production mechanism could thus be a continuous replenishment of PAHs from larger dust grains.  \citet{habart2004} suggested an evaporation of icy grain mantles within the disk. Since PAHs are ubiquitous in the ISM, they may have been included into ices during the cold molecular cloud phase. \citet{rafikov2006} suggested a PAH production through fragmentation of larger grains; a similar process has been discussed by \citet{berne2009} for the VSGs. This could fit into the more recent ideas that the dust grain size distribution in disks reflects the outcome of continuous coagulation and fragmentation processes \citep{birnstiel2010}.

\subsection{Do PAHs shape the protoplanetary disk structure?}

Photoelectric heating of even small amounts of PAHs is generally more efficient in heating the gas than the rest of the entire dust population. As shown by many authors \citep[e.g.][]{kamp2004, jonkheid2004,nomura2005,gorti2008}, gas and dust temperature de-couple in the disk surfaces, above $A_V \sim 1$, and the higher gas temperature determines the vertical scale height and hence flaring of the disk. The recently developed disk modeling code {\sc ProDiMo} solves self-consistently for the gas chemistry, energy balance and disk structure \citep{woitke2009,kamp2010} also including PAH chemistry and physics (Woitke et al.\ in preparation). A recent study of {\sc ProDiMo} disk models around Herbig stars shows that reducing the fractional abundance of PAHs in disks leads in general to much flatter disk structures and even strong shadows (van der Plas et al. in preparation).

\subsection{How can we use PAHs as tracers of processes in protoplanetary disks?}

The ambiguity whether PAHs are very small dust grains or large gas molecules makes them universal tracers of both disks components and widens their diagnostic potential.

As small dust grains, their emission can tell us if and where such small solid particles survive the settling and coagulation processes that lead otherwise to a rapid grain growth to mm- and cm-sized dust \citep[e.g.][]{natta2004}. PAHs can also be used to trace the radial disk structure up to $\sim 100$~AU. Since these grains are stochastically heated, their emission features are generally more extended than that of the continuum. As an example, \citet{geers2007a} imaged IRS48 (previously classified as M dwarf and recently re-classified as A dwarf) in the 8.6, 9, 11.3, 11.9, 18.7~$\mu$m filters. The 18.7~$\mu$m image shows a gap with a radius of 30~AU, while the shorter PAH bands are all centered on source. Hence, they conclude that the gap cannot be entirely devoid in dust. This object maybe caught in a short lived phase very much related to transitional disks such as HD\,141569. Another example for deviations from a simple smooth disk structure observed in PAH band emission are presented in \citet{doucet2005} for the Herbig star HD\,97048.

As large molecules, PAHs are thought to remain well mixed with the gas in the disk. As such, the PAH emission can also probe the physical conditions in the disk surfaces such as flaring and  irradiation. VISIR imaging of the disk around HD97048 in the PAH bands has shown that the observed emission, flaring angle and vertical scale height are consistent with the predictions from passive irradiated hydrostatic equilibrium disk models \citep{lagage2006}. This gives support to the idea that PAHs rather belong to the gas than to the dust component of disks.

Along the same lines, \citet{grady2005} show for a sample of 14 Herbig disks that the visibility from STIS coronagraphic imaging (sensitive to 0.5"-15" distances from star, hence $r \geq 50-70$~AU) is correlated with the strength of PAH features (particularly the 6.2~$\mu$m band). They report a correlation with disk flaring, and an anticorrelation with dust settling and the absence of any correlations with SED type, far-IR slope, mass accretion rate or strength of H$_2$ emission. This could suggest that there is initially a simple stratification with size, very small grains such as PAHs staying up high in the disk. During the disk evolution these small grains should be photodestroyed on short timescales, leaving a rather flat structure behind since photoelectric heating on large grains is generally less efficient. \citet{grady2005} also suggest that settling first occurs in the inner disk, so we should observe some disks that are flat inside and still flaring outside (if sufficient sample size available - since this might be a very short evolutionary phase), e.g.\ HD163296, DM\,Tau (faint ring with dark inner disk), which are both around 5 Myr. HD\,169142 ($\sim 6$~Myr) is an example, where the inner disk maybe de-coupled from outer disk  \citep{grady2007}. The spatially resolved PAH emission and the Meeus group\,I SED classification are only consistent if the inner disk is substantially flatter than the outer disk. 

PAHs are also good diagnostics of the stellar radiation field as the relative band strength indicate the fractional charging and typical sizes of PAHs which is directly related to the strength of the stellar radiation field. The stronger the radiation field, the less small PAHs can survive \citep[e.g.][]{visser2007}. This seems to be also confirmed by an analysis of the mid-IR emission of a sample of 12 {\it Spitzer} sources spread over the spectral ranges F to B \citep{berne2009}.

\paragraph{Acknowledgement}
This work made use of the {\sc ProDiMo} code developed by P.\ Woitke, W.-F.\ Thi and I.\ Kamp.

\bibliography{references}
\end{document}